%% file: main.tex
\documentclass[sigconf]{acmart}
\usepackage{listings}
\usepackage{xcolor}
\usepackage{enumitem}
\usepackage{algorithm}
\usepackage{algpseudocode}
\usepackage{threeparttable,booktabs}
\usepackage{array}
\usepackage{makecell}
\usepackage{cellspace}
\usepackage{colortbl}
\usepackage{graphicx}
\usepackage{multirow}
\usepackage{textcomp}
\usepackage{tcolorbox}
\usepackage{todonotes}
\usepackage{geometry}
\usepackage[utf8]{inputenc}

\usepackage[british]{babel}

\definecolor{eclipseStrings}{RGB}{42,0,255}
\definecolor{eclipseKeywords}{RGB}{127,0,85}
\definecolor{bgcolor}{HTML}{f5f4f2}
\colorlet{numb}{magenta!60!black}

\lstdefinelanguage{json}{
    basicstyle=\small\ttfamily,
    commentstyle=\color{eclipseStrings},
    stringstyle=\color{eclipseKeywords},
    numbers=left,
    numberstyle=\ttfamily,
    stepnumber=1,
    numbersep=8pt,
    showstringspaces=false,
    breaklines=true,
    backgroundcolor=\color{bgcolor},
    string=[s]{"}{"},
    comment=[l]{:\ "},
    morecomment=[l]{:"},
    literate=
        *{0}{{{\color{numb}0}}}{1}
         {1}{{{\color{numb}1}}}{1}
         {2}{{{\color{numb}2}}}{1}
         {3}{{{\color{numb}3}}}{1}
         {4}{{{\color{numb}4}}}{1}
         {5}{{{\color{numb}5}}}{1}
         {6}{{{\color{numb}6}}}{1}
         {7}{{{\color{numb}7}}}{1}
         {8}{{{\color{numb}8}}}{1}
         {9}{{{\color{numb}9}}}{1}
}

\makeatletter
\newcommand{\TODO}[1]{%
  \bgroup
  \def\@tempa{#1}%
  \expandafter\textcolor\expandafter{red}{\@tempa}%
  \GenericWarning{}{LaTeX Warning: TODO: \@tempa}%
  \egroup
}
\makeatother

\makeatletter
\newcommand{\NOTE}[1]{%
  \bgroup
  \def\@tempa{#1}%
  \expandafter\textcolor\expandafter{blue}{\@tempa}%
  \GenericWarning{}{LaTeX Warning: NOTE: \@tempa}%
  \egroup
}
\makeatother

\AtBeginDocument{
  }
\copyrightyear{2024}
\acmYear{2024}
\setcopyright{rightsretained}
\acmConference[ASE '24]{39th IEEE/ACM International Conference on Automated Software Engineering }{October 27-November 1, 2024}{Sacramento, CA, USA}
\acmBooktitle{39th IEEE/ACM International Conference on Automated Software Engineering (ASE '24), October 27-November 1, 2024, Sacramento, CA, USA}
\acmDOI{10.1145/3691620.3695356}
\acmISBN{979-8-4007-1248-7/24/10}

\begin{document}
\title{HighGuard: Cross-Chain Business Logic Monitoring of Smart Contracts}

\author{Mojtaba Eshghie}
\affiliation{%
  \institution{KTH Royal Institute of Technology}
  \city{Stockholm}
  \country{Sweden}
}
\email{eshghie@kth.se}

\author{Cyrille Artho}
\affiliation{%
  \institution{KTH Royal Institute of Technology}
  \city{Stockholm}
  \country{Sweden}
}
\email{artho@kth.se}

\author{Hans Stammler}
\affiliation{%
  \institution{KTH Royal Institute of Technology}
  \city{Stockholm}
  \country{Sweden}
}
\email{stammler@kth.se}

\author{Wolfgang Ahrendt}
\affiliation{%
  \institution{Chalmers University of Technology}
  \city{Gothenburg}
  \country{Sweden}
}
\email{ahrendt@chalmers.se}

\author{Thomas T. Hildebrandt}
\affiliation{%
  \institution{University of Copenhagen}
  \city{Copenhagen}
  \country{Denmark}
}
\email{hilde@di.ku.dk}

\author{Gerardo Schneider}
\affiliation{%
  \institution{University of Gothenburg}
  \city{Gothenburg}
  \country{Sweden}
}
\email{gerardo.schneider@gu.se}

\renewcommand{\shortauthors}{Eshghie et al.}

\begin{abstract}
Logical flaws in smart contracts are often exploited, leading to significant financial losses. Our tool, HighGuard, detects transactions that violate business logic specifications  of smart contracts. HighGuard employs dynamic condition response (DCR) graph models as formal specifications to verify contract execution against these models. It is capable of operating in a cross-chain environment for detecting business logic flaws across different blockchain platforms. We demonstrate HighGuard's effectiveness in identifying deviations from specified behaviors in smart contracts without requiring code instrumentation or incurring additional gas costs. By using precise specifications in the monitor, HighGuard achieves detection without false positives. Our evaluation, involving $54$ exploits, confirms HighGuard's effectiveness in detecting business logic vulnerabilities.

Our open-source implementation of HighGuard and a screencast of its usage are available at: \\
\href{https://github.com/mojtaba-eshghie/HighGuard}{https://github.com/mojtaba-eshghie/HighGuard}\\
\href{https://www.youtube.com/watch?v=sZYVV-slDaY}{https://www.youtube.com/watch?v=sZYVV-slDaY}
\end{abstract}



\maketitle

\input{macros}

\section{Introduction}
Smart contracts are computer programs that execute on blockchain platforms and manage digital assets. Smart contracts operate autonomously according to a predefined set of rules implemented in high-level programming languages such as Solidity~\cite{solidity-web}. They embody complex business processes, which in lack of business process-oriented development languages may lead to implementations that deviate from the intended business logic of the contract. Such flaws enable attackers to exploit the contracts~\cite{SunWeb3SecDeFiHackLabsReproduce}.

Popular programming lanaguages for smart contracts, such as Solidity, lack explicit support for process-oriented concepts such as roles, action dependencies, and time. This makes it difficult to design and analyze business logic directly in the smart contract. To address this problem, we use dynamic condition response (DCR) graphs to model smart contracts and their corresponding business processes \cite{captureDCR}. 
DCR graphs are a well-established declarative business process notation that extended with data and time provide a clear and concise model of the smart contract~\cite{dcrdata}.

\begin{figure*}
    \centering
    \includegraphics[width=.9\textwidth]{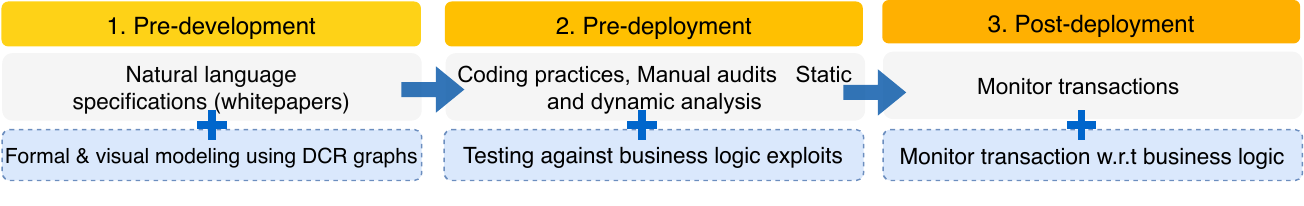}
    \caption{Software maintenance ecosystem of smart contracts (bottom blue row: HighGuard's approach)}
    \label{fig:sc-maintenance}
\end{figure*}

We leverage DCR graphs to express the intended design of a smart contract throughout its development cycle (see Fig.~\ref{fig:sc-maintenance}). The formal contract model helps convey the protocol designers' intentions to developers (stage one in Fig.~\ref{fig:sc-maintenance}) and supports later development and maintenance stages (stages two and three in Fig.~\ref{fig:sc-maintenance}).

Business logic flaws in smart contracts account for a significant portion of the total losses in recent smart contract attacks~\cite{SunWeb3SecDeFiHackLabsReproduce}. Previous research has largely overlooked business-logic vulnerabilities, focusing instead on well-known issues like reentrancy and integer overflow~\cite{toolInsights2023,feist2019slither}. This is because the business logic of a smart contract is application-specific. Identifying business logic flaws requires understanding the contract’s intended behavior. These flaws are not easily recognizable as they deviate from the expected program behavior and do not follow traditional patterns, complicating their detection by (esp.\ static) analysis tools.

To address the gap in detection of business logic exploits, we introduce HighGuard, a runtime monitoring tool for smart contracts. 
It leverages DCR graph specifications as oracles to differentiate between intended behavior and interactions that violate a contract's intended business logic. DCR graphs have been established as a suitable formalism to capture the security properties in smart contracts~\cite{captureDCR}. 

To mitigate the performance overhead of runtime monitoring (additional \emph{gas} usage), we execute the monitor \emph{off-chain}. Nevertheless, HighGuard is an \emph{online} monitor, as it observes the transactions as they are appended to the blockchain in near real-time.

Thanks to its architecture, HighGuard can support multiple chain environments and even monitor the business logic of cross-chain transactions, making it (to our knowledge) the first tool that is capable of this.

% the following pargraph is no longer valid (we don't look merely at events anymore, and we are multi-chain now;
%It monitors the Ethereum network, capturing events emitted by the contract and verifying their adherence to the contract's DCR graph specification. Should any deviation from the specification be detected, HighGuard generates an alert. Our application of HighGuard reveals that HighGuard effectively identifies deviations from the contract DCR specifications.

\section{Related Work}\label{sec:related-works}

\begin{table}[!t]
\scriptsize
\begin{center}
    \caption{SotA Smart contract monitoring tools}
    \label{tab:toolComparison}
    \renewcommand{\arraystretch}{1.5}
    \begin{tabular}{lm{0.95cm}m{1.25cm}m{1.8cm}m{0cm}} 
        \textbf{Tool} & \centering{\textbf{Monitor\\Placement}} & \centering\textbf{Evaluation\\Dataset Size} & \centering\textbf{Target \\Vulnerabilities} & \\
        \hline
        ContractLarva~\cite{gordonpace_gordonpacecontractlarva_2024} & On-chain & 1 contract~\cite{ellul_runtime_2018} & --- \\
        \hline
        Solythesis~\cite{li_aoli-solythesis_2024} & On-chain & 23 contr.~\cite{li_securing_2020} & --- \\
        \hline
        ContraMaster~\cite{noauthor_ntu-srslabvultron_2023} & Instr.\ EVM & 218 contr.~\cite{wang_oracle-supported_2022} & \makecell[ml]{Reentrancy\\Exception Disorder\\Integer Over/underflow} \\
        \hline
        Dynamit~\cite{dynamitGithub} & Off-chain & 105 tx~\cite{dynamicVul} & Reentrancy\\
        \hline
        SCMon~\cite{ding_function-level_2020} & Instr.\ EVM & 1 contract~\cite{ding_function-level_2020} & --- \\
        \hline
        Xscope~\cite{zhang_xscope_2023} & Off-chain & \makecell[l]{4 cross-chain\\bridges~\cite{xscope-tool_xscope-toolresults_2024}} & \makecell[ml]{Unrestricted Deposit \\ Inconsistent Event Parsing \\ Unauthorized Unlocking} \\
        \hline
        Annotation~\cite{shyamasundar_framework_2022} & On-chain & 50 contr.~\cite{shyamasundar_framework_2022} & \makecell[ml]{Reentrancy, Type cast,\\Tx order non-determinism\\Exception disorder} \\
        \hline
        Scribble~\cite{noauthor_consensysscribble_2024} & On-chain & --- & --- \\
        \hline
        {Tx Monitors~\cite{capretto_transaction_2022}} & Instr.\ EVM & --- & --- \\
        \hline
        Forta~\cite{noauthorforta2022} & Off-chain & --- & --- \\
        \hline
        HAL Streams~\cite{noauthorhalnodate} & Off-chain & --- & --- \\
        \hline
        OpenZeppelin~\cite{OZMonitor} & Off-chain & --- & --- \\
        \hline
        \textbf{HighGuard} & Off-chain & $54$ exploits & Business logic flaws \\
    \end{tabular}
  \end{center}
\end{table}

Smart contract analysis tools include static and dynamic methods. Table~\ref{tab:toolComparison} summarizes state-of-the-art monitoring tools.

\paragraph{Static Analysis.} Tools like Slither \cite{feist_slither_2023}, SmartCheck \cite{tikhomirov_smartcheck_2018}, and Securify \cite{tsankov_securify_2018}, identify patterns in code but often produce false positives due to syntax-level checks \cite{shyamasundar_framework_2022, wang_oracle-supported_2022}.

\paragraph{Dynamic Analysis.} Dynamic analysis tools vary in their monitoring approach and target vulnerabilities. ContractLarva adds runtime checks based on automaton-based specifications \cite{gordonpace_gordonpacecontractlarva_2024, ellul_runtime_2018}, while Solythesis enforces invariants through a source-to-source compiler \cite{li_aoli-solythesis_2024, li_securing_2020}. Shyamasundar's framework allows in-code constraints enforced through safeguards \cite{shyamasundar_framework_2022}. Scribble generates runtime assertions from annotations for pre-deployment testing \cite{noauthor_introduction_2023}. Capretto et al. propose transaction monitors validating transaction conditions, requiring new blockchain instructions \cite{capretto_transaction_2022}.

Pre-deployment tools like ContraMaster use grey-box fuzzing to test attack transactions \cite{noauthor_ntu-srslabvultron_2023, wang_oracle-supported_2022}. SCMon logs and visualizes function-level execution metrics~\cite{ding_function-level_2020}.

Post-deployment tools monitor deployed contracts for malicious traces. Dynamit uses machine learning to classify transactions, focusing on reentrancy attacks~\cite{dynamicVul}. Forta employs decentralized detection bots~\cite{noauthorforta2022,noauthornetworknodate}. HAL Streams filters blockchain data for specific events~\cite{noauthorhalnodate}. OpenZeppelin Monitors provide alerts for specific events~\cite{OZMonitor,noauthordefendernodate}.

\paragraph{Chain Interoperability.} Centralized bridges enable asset transfer between blockchains~\cite{buterin_chain_2016,augusto_sok_2024,ou_overview_2022}. Xscope identifies vulnerabilities by pre-executing transaction sequences as a relayer in cross-chain bridges~\cite{zhang_xscope_2023}. Ganguly et al. propose distributed runtime verification across blockchains~\cite{ganguly_distributed_2022}. Unlike these tools, HighGuard supports both pre- and post-deployment testing and monitoring through its multi-chain execution ecosystem (stages two and three in Fig.~\ref{fig:sc-maintenance}). Rather than using predefined anomalous transaction sequences, HighGuard relies on the reference DCR model of the contract as the monitoring oracle (Table~\ref{tab:toolComparison}, last row), enabling developers to apply various contract-specific high-level properties to the monitor.

\section{HighGuard Architecture}\label{sec:arch}

\begin{figure}%[!t]
    \centering
    \includegraphics[width=0.45\textwidth]{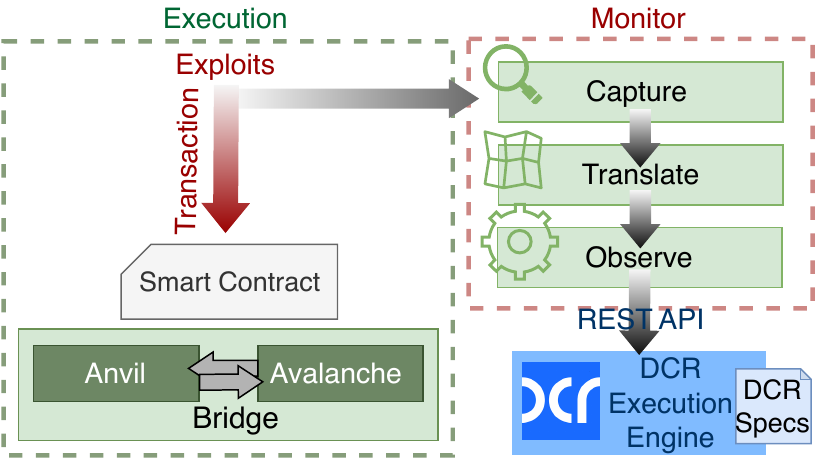}
    \caption{HighGuard system architecture}
    \label{fig:highguard-arch}
\end{figure}

HighGuard requires two inputs from the user: (1) a DCR model of the contract, and (2) the mapping of contract functions, events, and transaction(s) to DCR model activities.  

Fig.~\ref{fig:highguard-arch} shows HighGuard's architecture. The system monitors incoming transactions to smart contracts and translates them to DCR graph activities based on transaction information. The monitor then sends requests based on the translated events to the DCR execution engine.
The trace of executed activities is generated as a report while the monitor is running.
If the event is part of a violating trace in the DCR model, it will return an error code that is logged in the monitor for further investigation.

\begin{figure}
    \centering
    \includegraphics[width=0.4\textwidth]{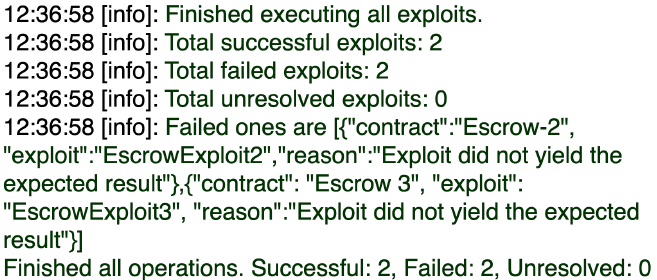}
    \caption{CLI Termination output for Escrow contract.}
    \label{fig:cli-output}
\end{figure}

\begin{figure}
    \centering
    \includegraphics[width=0.42\textwidth]{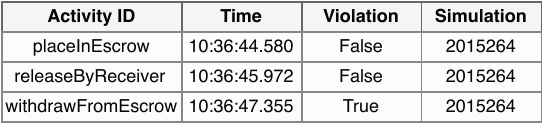}
    \caption{Report generated for Escrow contract}
    \label{fig:escrow-report}
\end{figure}

HighGuard offers a command line interface to deploy, run, and report. Fig.~\ref{fig:cli-output} shows an example report produced by executing four exploits against four vulnerable Escrow contracts.
Fig.~\ref{fig:escrow-report} shows details of a particular exploit.

%Fig.~\ref{fig:highguard-arch} depicts system design of HighGuard. 
HighGuard is intended for two types of usage: (1) In a pre-deployment testing setup (stage two in Fig.~\ref{fig:sc-maintenance}) through the execution ecosystem of HighGuard (left side of Fig.~\ref{fig:highguard-arch}); (2)~Plugged into blockchain environments such as Ethereum to monitor contracts post-deployment (stage three in Fig.~\ref{fig:sc-maintenance}).

HighGuard supports multiple blockchain simulators as \emph{environments} that abstract away details of each simulation platform and expose an API to the monitor for deploying contracts, executing exploits, and monitoring for violations from exploits. Two such blockchain simulators, \emph{Anvil} and \emph{Avalanche}, are implemented as environments~\cite{FoundryBook,AvalancheCLIAvalancheDev}.
%\todo[inline]{Left side of Fig. 1: stages 1 and 2? Clarify. Also cover, "the latter" part. Further, the description of environments/plugins seems unrelated to the usage within the life cycle?}

HighGuard is capable of monitoring cross-chain transactions as the business logic specified in the DCR model of the contract is platform-independent~\cite{captureDCR}. To execute cross-chain transactions, we implement a bridge that supports cross-chain transactions between the two aforementioned environments (left side of Fig.~\ref{fig:highguard-arch}).\footnote{\url{https://github.com/mojtaba-eshghie/HighGuard/tree/main/CI/envs/bridge-decentralized}}
%This ensures that any deviations from the specified behaviors are detected without modifying the smart contract code or affecting its execution on the blockchain.

\section{DCR Graph Modeling}
\begin{figure}
    \centering
    \includegraphics[width=0.49\textwidth]{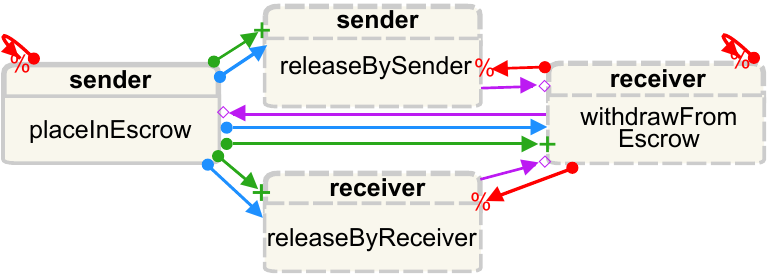}
    \caption{DCR model of the Escrow contract.}
    \label{fig:escrow-dcr}
\end{figure}

DCR graphs contain activities (boxes in Fig.~\ref{fig:escrow-dcr}) and relations between activities (arrows in Fig.~\ref{fig:escrow-dcr}). 
Smart contracts implement functionality as functions that affect state variables. 

To model the semantics of contracts in DCR graphs, we represent publicly callable functions of a contract as activities in the DCR graph. The condition checks (\emph{require} statements in Solidity) that preserve invariants in the contract %for function executions 
are mapped to the relations in DCR graphs. A function's requirement in the form of \emph{require(predicate)} in Solidity is translated to a guarded inclusion (\includerel) or exclusion (\excluderel) relation~\cite{dcrdata} in the DCR graph with \emph{predicate} written as a guard on top of the relation in the model.

A DCR graph activity is enabled if it is included (drawn with solid border, e.\,g., \emph{placeInEscrow} in Fig.~\ref{fig:escrow-dcr}). % and all other activities that are sources of a condition relation pointing to it are either excluded (drawn with a dashed border) or has been executed.
%where the enabling of the target event %depends on the execution or exclusion of %the source event. Similarly,
\begin{comment}
State changes in Solidity, often managed through enumerated states, can be represented by inclusion (\includerel) and exclusion (\excluderel) relations in DCR graphs, e.g. when \texttt{applyDiscount} is executed, it excludes itself and includes the \texttt{payForOrder} activity. %which dictate 
%the partial ordering and enabling %conditions of events.
Time constraints are represented as DCR response (\responserel) relations with deadlines, e.g. in Fig.~\ref{fig:escrow-dcr} \texttt{payForOrder} must happen within $5$ seconds after \texttt{applyDiscount} is executed. 
\end{comment}
Time constraints are represented as DCR response (\responserel) relations with deadlines; inter-action dependencies can be modeled as milestones (\milestonerel) that govern when actions are enabled~\cite{dcrdata}.

Finally,
Solidity roles
are directly mapped to roles in DCR models (\emph{sender} and \emph{receiver} in Fig.~\ref{fig:escrow-dcr}), ensuring that each event is executed only by the permitted actors.

\section{Evaluating HighGuard}
The DCR models can be designed and simulated in the \url{dcrgraphs.net} online tool and then executed via a REST API.\footnote{The online tool and API can be used freely for academic use.}
During runtime, a sequence of one or more transactions in the smart contract are translated to one activity execution in the DCR model depending on the mapping given to HighGuard for each contract.

\paragraph{Single-Chain Evaluation.} We modeled five contracts using DCR graphs, including the example in Fig.~\ref{fig:escrow-dcr}. Variants of each contract were deployed with business logic vulnerabilities injected into their Solidity source code, such as altered equations in functions and removed \emph{require} statements. One author reviewed these vulnerabilities to exclude traditional types like reentrancy, focusing instead on deviations from the contract's business specifications.\footnote{\href{https://github.com/mojtaba-eshghie/HighGuard/tree/main/contracts/src/synthesized}{https://github.com/mojtaba-eshghie/HighGuard/tree/main/contracts/src/synthesized}} We evaluated HighGuard's ability to detect malicious transactions by running exploits~\footnote{\href{https://github.com/mojtaba-eshghie/HighGuard/tree/main/CI/exploits/synthesized}{https://github.com/mojtaba-eshghie/HighGuard/tree/main/CI/exploits/synthesized}} targeting these vulnerabilities using HighGuard's ecosystem (Fig.~\ref{fig:highguard-arch}). In total, 52 pairs of vulnerable contract variants and their exploits were tested. Table~\ref{tab:results} shows the detection results: HighGuard flagged all exploits with no false positives or false negatives. This experiment was conducted in the \emph{Anvil} environment (Fig.~\ref{fig:highguard-arch}).

\paragraph{Cross-Chain Evaluation.}
We modeled a cross-chain decentralized exchange (DEX) with four contracts: a vault, a token, a price oracle, and a router contract on each of Ethereum and Avalanche blockchains. As mentioned earlier, the execution ecosystem uses our own centralized bridge for pre-deployment testing purposes (Fig.~\ref{fig:highguard-arch}). 
%\NOTE{briefly talk about ecapabilities of this DEX (esp. cross-chain ones)}
Our cross-chain DEX contract can exchange tokens between two blockchains using the exchange rate updated by the price oracle contract.
%\NOTE{briefly introduce the vulnerabilities injected}
We instrumented the implementations of the \emph{vault} and \emph{router} contracts to inject two vulnerabilities related to cross-chain transaction expiration times and double-payouts.  
We ran manually-written exploits\footnote{\href{https://github.com/mojtaba-eshghie/HighGuard/tree/main/CI/tests}{https://github.com/mojtaba-eshghie/HighGuard/tree/main/CI/tests}} targeting the mentioned vulnerabilities which resulted in loss of tokens in victim contracts. These two exploits were successfully detected by HighGuard with no false positives or false negatives.

\paragraph{Resource Usage.}

Since the off-chain monitor placement does not affect contracts under observation, there is no on-chain performance overhead. The monitor application, written in NodeJS, runs on a server with memory usage under $1\,\mathrm{GB}$. In a pre-deployment testing setup, HighGuard also provides the smart contract execution platform, with resource usage depending on the testing environment. In the \emph{Anvil} environment, it uses less than $10\,\mathrm{MB}$ of RAM and 1\% CPU on a MacBook with an Intel i7 $2.30\,\mathrm{GHz}$ processor.
%the resource usage in Fig.~\ref{fig:resource-usage-anvil} when using \emph{Anvil} as the simulated blockchain for deploying one contract and running one exploit with $6$ transactions.

\begin{comment}
    \begin{figure}
    \centering
    \includegraphics[width=0.4\textwidth]{figures/perf.pdf}
    \caption{Resource usage of Anvil}
    \label{fig:resource-usage-anvil}
\end{figure}
\end{comment}

% \begin{table}
%   \caption{Evaluation Results}
%   \label{tab:results}
%   \footnotesize
%   \begin{tabular}{lccc}
%     %\toprule
%     \textbf{Contract} & \textbf{Exploitable} & \textbf{Partly Exploitable} & \textbf{FP/FN} \\
%     \midrule
%     Governance & 4 & 11 & 0/0 \\
%     Escrow & 1 & 1 & 0/0 \\
%     MultiStageAuction & 0 & 13 & 0/0 \\
%     PrizeDistribution & 0 & 7 & 0/0 \\
%     ProductOrder & 10 & 5 & 0/0 \\
%     \midrule
%     Total & 15 & 37 & 0/0 \\
%   %\bottomrule
% \end{tabular}
% \end{table}

\begin{table}
  \caption{Single-chain evaluation results}
  \label{tab:results}
  \small
  \begin{tabular}{lccc}
    %\toprule
    \textbf{Contract} & \textbf{Exploits} & \textbf{FP} & \textbf{FN} \\
    \midrule
    Governance & 15 & 0 & 0 \\
    Escrow & \hphantom{0}2 & 0 & 0 \\
    MultiStageAuction &  13 & 0 & 0 \\
    PrizeDistribution &  \hphantom{0}7 & 0 & 0 \\
    ProductOrder &  15 & 0 & 0 \\
    \midrule
    Total & 52 & 0 & 0 \\
  %\bottomrule
\end{tabular}
\end{table}

\section{Conclusion}
We present HighGuard, a tool for detecting business logic flaws in smart contracts.
HighGuard takes the DCR model of a contract's business logic as the reference to check the transactions against it, pre- or post-deployment of the contract. It operates off-chain and does not require any code instrumentation. We successfully demonstrated HighGuard's capability of detecting malicious transactions in single-chain and cross-chain setups by evaluating it against $54$ smart contract exploits. 
%Its ability to operate off-chain without code instrumentation and gas costs makes it an effective solution. 

\bibliographystyle{ACM-Reference-Format}
\bibliography{refs}

\end{document}

%% file: macros.tex
\definecolor{responseColor}{HTML}{3888D6}
\definecolor{conditioncolor}{HTML}{CC7F0C}
\definecolor{milestoneColor}{HTML}{BA1FE5}
\definecolor{includeColor}{HTML}{2FA71F}
\definecolor{excludeColor}{HTML}{C10300}
\definecolor{noresponseColor}{HTML}{8c6026} 
\definecolor{valueColor}{HTML}{8c8c8c} 

\newcommand{\evalexp}[2]{\ensuremath{[[#1]]_{#2}}}
\newcommand{\DCR}{DCR\xspace}
\newcommand{\DCRR}{DCR$^*$\xspace}
\newcommand{\DCRL}{DCR$^\nu$\xspace}
\newcommand{\DCRB}{DCR$^!$\xspace}
\newcommand{\effectoff}[2]{\ensuremath{#1\cdot #2}}
\newcommand{\inputact}[2]{\ensuremath{?(#1, #2)}}
\newcommand{\outputact}[3]{\ensuremath{!(#1, #2, #3)}}
\newcommand{\outact}[2]{\ensuremath{!(#1, #2)}}
\newcommand{\intact}[2]{\ensuremath{*(#1, #2)}}
\newcommand{\ioactions}{\ensuremath{\mathsf{IO_{A,D}}}}
\newcommand{\inoutactions}{\ensuremath{\mathsf{IO_{A}}}}
\newcommand{\interact}[3]{\ensuremath{(#1, #2 \rightarrow #3)}}
\newcommand{\actt}[3]{\ensuremath{#1(#2,#3)}}
\newcommand{\interactdata}[4]{\ensuremath{(#1, #2 \stackrel{#3}{\longrightarrow} #4)}}
\newcommand{\rref}{\sqsubseteq}
\renewcommand\t{\ensuremath{\mathsf{t}}}
\newcommand\f{\ensuremath{\mathsf{f}}}
\newcommand{\VAL}{\ensuremath{V}} 
\newcommand{\natinf}{\ensuremath{\infty}}

\newcommand{\lab}{\ensuremath{l}}
\newcommand{\valuerel}{{\color{valueColor} \ensuremath{\mathrel{\rightarrow\!\!=}}}}
\newcommand{\conditionrel}{{\color{conditioncolor} \ensuremath{\mathrel{\rightarrow\!\!\bullet}}}}
\newcommand{\responserel}{{\color{responseColor}\ensuremath{\mathrel{\bullet\!\!\rightarrow}}}}
\newcommand{\milestonerel}{{\color{milestoneColor} \ensuremath{\mathrel{\rightarrow\!\!\diamond}}}}
\newcommand{\includerel}{{\color{includeColor} \ensuremath{\mathrel{\rightarrow\!\!\textsf{+}}}}}
\newcommand{\excluderel}{{\color{excludeColor} \ensuremath{\mathrel{\rightarrow\!\!\textsf{\%}}}}}

\newcommand{\noresponserel}{{\color{noresponseColor}\ensuremath{\mathrel{\bullet\!\!\!\rightarrow\!\!\!\times}}}}

\newcommand{\gconditionrel}[2]{{\color{conditioncolor}\ensuremath{\stackrel{[#1]}{\mathrel{\rightarrow\!\!\bullet}}_{d}}}}
\newcommand{\gresponserel}[2]{{\color{responseColor}\ensuremath{\stackrel{[#1]}{\mathrel{\bullet\!\!\rightarrow}}_{#2}}}}
\newcommand{\gmilestonerel}[1]{{\color{milestoneColor}\ensuremath{\stackrel{[#1]}{\mathrel{\rightarrow\!\!\diamond}}}}}
\newcommand{\gincluderel}[1]{{\color{includeColor}\ensuremath{\stackrel{[#1]}{\mathrel{\rightarrow\!\!+}}}}}
\newcommand{\gexcluderel}[1]{{\color{excludeColor}\ensuremath{\stackrel{[#1]}{\mathrel{\rightarrow\!\!\%}}}}}
\newcommand{\gnoresponserel}[1]{{\color{noresponseColor}\ensuremath{\stackrel{[#1]}{\mathrel{\bullet\!\!\!\rightarrow\!\!\!\times }}}}}

\newcommand{\responses}{\ensuremath{\mathsf{Re}}}
\newcommand{\executed}{\ensuremath{\mathsf{Ex}}}
\newcommand{\included}{\ensuremath{\mathsf{In}}}

\newcommand{\markingset}{\ensuremath{\mathcal{M}}}
\newcommand{\graphs}{\ensuremath{\mathcal{G}}}
\newcommand{\pgraphs}{\ensuremath{{\cal{P}}}}

\newcommand{\genrel}{\ensuremath{\mathrel{\rightarrow}}}
\newcommand{\genrelto}[1]{\genrel\!#1}
\newcommand{\genrelfrom}[1]{#1\!\genrel}
\newcommand{\enable}[2]{\ensuremath{\mathsf{enabled}(#1,#2)}} 
\newcommand{\power}[1]{\ensuremath{{\cal P}(#1)}}
\newcommand{\maxc}{\ensuremath{maxc_G}}
\newcommand{\minr}{\ensuremath{minr_G}}
\newcommand{\miner}{\ensuremath{minRe_G}}
\def\L{\mathsf{L}}

\def\lEx{\mathsf{L_{Ex}}}
\def\lRe{\mathsf{L_{Re}}}
\def\lIn{\mathsf{L_{In}}}
\def\dom{\mathsf{dom}}

\newcommand{\spg}{\ensuremath{sp}} 
\newcommand{\ESG}{\ensuremath{E}}

\newcommand\Ex{\mathsf{Ex}}
\renewcommand\Re{\mathsf{Re}}
\newcommand\In{\mathsf{In}}

\newcommand\Va{\mathsf{Va}}

\newcommand{\commitevent}{\ensuremath{\mathsf{commit}}}
\newcommand{\revealevent}{\ensuremath{\mathsf{reveal}}}
\newcommand{\transactionevent}{\ensuremath{\mathsf{revealtransaction}}}
\newcommand{\placeInEscrowEvent}{\ensuremath{\mathsf{placeInEscrow}}}
\newcommand{\releaseBySenderEvent}{\ensuremath{\mathsf{releaseBySender}}}
\newcommand{\releaseByReceiverEvent}{\ensuremath{\mathsf{releaseByReceiver}}}
\newcommand{\withdrawFromEscrowEvent}{\ensuremath{\mathsf{withdrawFromEscrow}}}

%%%%%%%%%%%%%%%%%%%%%%%
\newcommand{\draftcomment}[3]{{\color{#1}[#3] #2} 
  \PackageWarning{WARNING: Draft comments visible}{#2: #3}}
\newcommand{\gs}[1]{\draftcomment{red}{GS}{#1} }
\newcommand{\wa}[1]{\draftcomment{blue}{WA}{#1} }
%%%%%%%%%%%%%%%%%%%%%%%